\documentclass[preprint,amsmath,amssymb]{revtex4}

\usepackage{graphicx}

\begin{document}
\title{The Interpretation of Wave Mechanics with the help of Waves with Singular Regions\footnote{This paper, whose original title was ``\emph{L'Interpr\'{e}tation de M\'{e}canique Ondulatoire \`{a} l'Aide d'Ondes \`{a} R\'{e}gions Singuli\`{e}res}",  has been translated from the French by
Dileep Karanth, Department of Physics, University of Wisconsin-Parkside, Kenosha, USA.}}
\author{Louis de Broglie}
\affiliation{ University of Paris}

\begin{abstract}
This paper appeared in a collection of papers titled ``Scientific Papers Presented to Max Born on his
retirement from the Tait Chair of Natural Philosophy  in the
University of Edinburgh'', published in 1953 (Oliver
and Boyd).
\end{abstract}


\maketitle

The homage paid to the great theoretical physicist that is Max
Born  highlights the role he has played in contemporary physics by
introducing the probabilistic interpretation of the wave $\psi$ of
wave mechanics. In the article he has written for this
felicitation volume, Einstein has summarized some of his
objections to the adoption of the ``purely probabilistic"
interpretation of quantum mechanics which has developed out of the
works of thinkers such as Born, Bohr and Heisenberg. I wish to
recall briefly what my ideas on this question formerly were and
why I have recently undertaken a fresh examination of these old
ideas.

Between 1924, when my doctoral thesis was published, and 1927, I
have tried to develop a causal and objective interpretation of
wave mechanics by admitting the hypothesis of ``double solution"
according to which the linear equations of wave mechanics allow
two kinds of solutions: the continuous solutions $\psi$ usually
considered, whose statistical nature was then beginning to emerge
clearly thanks to the work of Born, and singular solutions which
would have a concrete meaning and which would be the true physical
representation of particles. The latter would be well localized in
agreement with the classical picture, but would be incorporated in
an extended wave phenomenon. For this reason, the particle's
movement would not follow the laws of classical mechanics
according to which the particle is subject only to the action of
forces which act on it along its trajectory, and does not suffer
any repercussion from the existence of obstacles which may be
situated far away outside its trajectory: In my present
conception, on the contrary, the movement of the singularity
should experience the influence of all the obstacles which hinder
the propagation of the wave with which it is connected. This
circumstance would explain the existence of the phenomena of
interference and diffraction \cite{DeBroglie2}.

However, the development of the theory of the double solution
presented great mathematical difficulties. For this reason, I
contented myself with a simplified form of my ideas, to which I
gave the name ``pilot-wave theory", and which coincided with the
hydrodynamic interpretation of wave mechanics proposed at about
the same time by Madelung \cite{DeBroglie3}. I have presented this
softened form of my ideas at the October 1927 Solvay Physics
Conference. My presentation was the object of numerous criticisms
notably on the part of Pauli. Pauli's objections did not appear to
me as being decisive, but soon thereafter I recognized that the
pilot-wave theory was faced with a difficulty which seemed and
still seems to me to be insurmountable.

The wave $\psi$ used in wave mechanics cannot be a physical
reality: its normalization is arbitrary, its propagation is
supposed to take place in general in a visibly fictive
configuration space, and according to Born's ideas, it is only a
representation of probability depending on the state of our
knowledge and is suddenly modified by the information brought to
us by every new measurement. Thus, with only the help of the
pilot-wave theory, one cannot obtain a causal and objective
interpretation of wave mechanics, by supposing that the particle
is guided by the wave $\psi$. For this reason, since 1927 I had
entirely come round to the purely probabilistic interpretation of
Born, Bohr and Heisenberg.

A year and a half ago, David Bohm took up the pilot-wave theory
again. His work is very interesting in many ways and contains an
analysis of the measurement process which appears to be capable of
answering an objection Pauli had raised to me in 1927
\cite{Bohm1}. But since Bohm's theory regards the wave $\psi$ as a
physical reality, it seems to me to be unacceptable in its present
form. Reiterating the arguments I have recalled above, Takabayasi,
while demonstrating the interesting aspects of Bohm's ideas, has
recently insisted on the impossibility of admitting the principle
which is its point of departure \cite{Takabayasi}.
    But my first theory of the ``double solution" does not seem to me to hurtle against the same difficulties, because it distinguishes the singular wave, which alone is endowed with physical reality, from the wave $\psi$, which in this theory, is but a statistical representation of the relative probabilities of the possible movements of these singularities. J.P. Vigier having brought to my notice the analogies which existed between my considerations of 1927 and Einstein's ideas on the movement of particles considered as some kinds of field singularities in general relativity, I undertook once again the study of my old ideas about singular solutions.
    I will first recall the principles of my attempt of 1927. If $\psi = R
    e^{iS/\hbar}$ is a solution of the wave equation of the usual
    continuous type, we also admit the existence of solutions of
    the form:

\begin{equation}\label{1}
    u(x,y,z,t) = f(x,y,z,t)e^{\frac{iS(x,y,z,t)}{\hbar}}
\end{equation}

where $S(x,y,z,t)$ is the \emph{same} function as in $\psi$ and
where $f$ represents an in general mobile singularity. The
substitution of the solution $u$ into the wave equation leads to
the relation:

\begin{equation}\label{2}
    \frac{\partial f}{\partial t} + \frac{\partial f}{\partial n}
    \frac{1}{m}|\nabla S| + \frac{1}{2m}f\triangle S = 0
\end{equation}
at least in the non-relativistic case: the variable $n$ is
calculated along the normal to the surface $ S = constant$. It is
natural to suppose that, if the singularity is approached along
this normal, $f$ and $\frac{\partial f}{\partial n}$ are very
large and that $\frac{\partial f}{\partial n} \gg f$. Then, for
the velocity of the particle, $\vec{v}$, we find the fundamental
formula
\begin{equation}\label{3}
\vec{v} = \frac{\frac{\partial f}{\partial t}}{\frac{\partial
f}{\partial n}} = -\frac{1}{m}\nabla S
\end{equation}

Thus everything happens as \emph{if} the particle were guided by a
wave $\psi$, as in the pilot-wave theory revived by Bohm. However,
here, inasmuch as $S$ is also the phase of the wave $u$ which has
physical reality, the same objections do not arise.

Since the quantity $|\psi|^2 = R^2$ obeys the well-known
continuity equation
\begin{equation}\label{4}
\frac{\partial R^2}{\partial t} + \nabla \cdot (R^2 \vec{v}) = 0
\end{equation}
where $\vec{v}$ is given by equation $\ref{3}$. It is natural to suppose
that $R^2$ gives the probability of the presence of the
singularity at a point when it is not known which of its
trajectories is being described. Thus we come back to Born's
hypothesis about the statistical meaning of $|\psi|^2$. This
hypothesis appears here to be somewhat analogous to the one made
in statistical mechanics, when only on the basis of Liouville's
Theorem, one admits the equal probability of equal volumes of
phase space. But a more complete justification appears to be
necessary: in a recent memoir \cite{Bohm2}, Bohm has spelled out
an argument which seems to lead to this justification.

We may add that my theory of the double solution leads us, as I
have already shown in 1927, to consider the movement of the
particle as taking place under the action of classical forces
augmented by a quantum force derived from a potential:

\begin{equation}\label{5}
    U = -\frac{\hbar^2}{2m}\frac{\Box R}{R}= -\frac{\hbar^2}{2m}\frac{\Box f}{f}
\end{equation}

where the equality of the two expressions for $U$ follows from the
hypothesis of the equality of the phases of $\psi$ and $u$. The
potential $U$ is the ``quantum potential" of my 1927 theory
rediscovered by Bohm in his memoir.

It is evidently necessary to be able to extend the theory of the
double solution to the case of the Dirac equations for electrons
with spin. After a first attempt I made in this direction, Vigier
made a second which now seems preferable to me \cite{DeBroglie3}.
In any case, it does not seem to me that the extension of the
double solution to the Dirac equations raises essential
difficulties.

However it is necessary to demonstrate the existence of solutions
of the type $u$. Now an argument due principally to Sommerfeld
shows that in general in a problem involving quantized states
there do not exist singular solutions to the \emph{linear}
equations of wave mechanics having the same frequency as that of a
stationary wave $\psi$. This result proves that it is not possible
to consider the wave $u$ as a solution of these linear equations
possessing a singularity in the usual sense of the word, as I did
in 1927. But it is possible to overcome this objection by using
the term ``singularity" to mean a very small singular region where
the function $u$ takes values so large that the equation it
satisfies in this region is non-linear, with the linear equation
valid everywhere for $\psi$ being valid for $u$ only outside the
singular region \cite{DeBroglie5}. This change in point of view
does not alter the validity of the guiding formula (3), for which
we can give a proof more rigorous than the one sketched above.
Vigier thinks that one could thus reconcile the theory of the
double solution with the ideas of Einstein, who has always tried
to represent particles as singular regions of the field, and
probably also with the non-linear electromagnetism of Born.
Although it is not yet possible to pass definitive judgments on
Vigier's attempts, they allow us to entertain hopes of seeing the
theory of General Relativity and that of Quanta united within the
framework of a single representation in which causality will be
reestablished.

An important point would be to justify the use of the formula (3)
in the case of systems of interacting particles, $S$ then being
the phase of the wave $\psi$ in configuration space $\vec{v}$ the
velocity of the representative point in this space. It would be
necessary to show that this results from interactions between wave
singularities of the type $u$ evolving in three-dimensional
physical space. In my article in the May 1927 issue of the Journal
de Physique, I have sketched a proof of this kind, considering the
configuration space as being defined by the coordinates of the
singularities. In this way one is able to represent the movement
of the interacting singularities as taking place in the physical
space without necessarily making appeal to the configuration
space. This fictitious space and the propagation of the wave
$\psi$ in it become necessary only for statistical expectation
values. Following this line of argumentation, one should be able
to obtain a physical interpretation of the Pauli principle if it
could be shown that for fermions the wave $u$ can involve
 only one singularity, whereas it could have
several in the case of bosons. I have recently been able to
elaborate some considerations which I think constitute a slight
advancement in this direction \cite{DeBroglie4}.

The existence of singular regions of the wave $u$ (whose
dimensions most probably will be of the order of $10^{-13}$ cm)
may permit us to endow elementary particles with a structure whose
lack can be felt today in quantum theories, and probably even
resolve difficulties pertaining to infinite energies. As Born has
remarked, it is possible that in the atomic nucleus may be present
conditions not foreseen by current theories: in our way of
thinking, they will be due to an overlapping of the singular
regions of the constituents of the nucleus. One can also foresee
that the statistical role of the wave $\psi$ will not be valid
 in the case of particles animated by
movements so rapid that the length of their associated wave will
attain dimension comparable to those of their singular regions.

Before I conclude, I would like to say a word on the subject of
the objection raised by Einstein in his article against the
formula (3), an objection that applies both to the theory of the
double solution and to the pilot-wave theory. Like him, let us
consider a wave $\psi$ constrained to propagate along an axis
\emph{Ox} and to reflect off two perfectly reflecting mirrors
placed perpendicularly to the axis at $x=0$ and $x=l$. The
stationary forms of the wave $\psi$ are given by:
\begin{equation}\label{6}
    \psi_n = a_n sin \frac{n\pi x}{l}e^{\frac{iE_n t}{\hbar}}
\end{equation}

When one of these stationary states is realized, equation (3)
gives $\vec{v} = 0$. The particle associated with the stationary
wave is immobile. Now, says Einstein, this consequence of equation
(3) is inadmissible because it should be exact whatever the mass
of the particle, and if this particle has a macroscopic mass and
constitutes say a small ball, it is well known that its movement
should consist of a to and fro motion along the \emph{Ox} axis,
with alternate rebounds from each mirror. Without going into a
discussion of Einstein's very interesting argument in all its
generality, I will limit myself to the following remark. In order
for an expression such as (6) to be physically valid, it is
necessary that the plane surface of the two mirrors be well
defined at the scale of the wavelength. The mirrors are
necessarily made of atoms in thermal movement, and as a
consequence the precision with which the surfaces are defined
cannot be greater than a fraction of an angstr\"{o}m. The
condition
\begin{equation*}\label{nonumber}
    \lambda = \frac{h}{mv} > 10^{-9}
\end{equation*}
in c.g.s. units shows that if the particle has macroscopic mass
(say greater than $10^{-9}$ grams), the velocity $v$ should be
practically zero. In order for the expression (6) to be considered
as valid for a particle of macroscopic mass, its velocity must be
practically zero, and the value given by (3) is practically
satisfied.

In conclusion, I recognize that many difficulties still stand in
the way of the adoption of the theory of the double solution.
However, in spite of the risk of an ultimate failure, it does not
seem to me to be useless to take up again the ideas I have
recalled, to see whether, when suitably modified or completed,
they can provide a causal and objective interpretation of wave
mechanics, in accordance with the wish expressed many times by
Einstein. If some day such a thing were to come to pass, it would
of course hardly detract from the importance of the discovery made
by Born the day he apprehended the statistical meaning of $\psi$
in the usual wave mechanics.

\section*{Translator's Acknowledgments}
The translator is grateful to the Louis de Broglie Foundation for permission to translate this paper, and publish it (private communication, May 21, 2010).


\begin{thebibliography}{1}
\bibitem{Einstein} A. Einstein, this volume, p. 33.
  \bibitem{DeBroglie2} Louis de Broglie, \emph{Compt. Rend.} {\bf 183},
  447 (1926). Louis de Broglie, \emph{Compt. Rend.} {\bf 184},
  273 (1927). Louis de Broglie, \emph{Compt. Rend.} {\bf 185},
  380 (1927). Louis de Broglie, \emph{Journal de Physique}, series VI, t. 8, p. 225, (May 1927).
\bibitem{DeBroglie3} Louis de Broglie, \emph{Electrons et
 Photons}, Report to the Vth Solvay Conference, Gauthier Villars,
 p. 115 (1930). \emph{Introduction \`{a} l'\'{e}tude de la
 M\'{e}canique ondulatoire}, Hermann, Paris, 1930 (English edition,
 Methuen, London).
 \bibitem{Bohm1} David Bohm, \emph{Physical Review},{\bf 85}, 166
 and 180 (15th January 1952).
 \bibitem{Takabayasi} Takehiko Takabayasi, \emph{Progress of Theoretical
 Physics}, {\bf 8}, 143 (August 1952).
  \bibitem{Bohm2} David Bohm, \emph{Physical Review},{\bf 89}, 458
(15th January 1953).
 \bibitem{DeBroglie3} Louis de Broglie, \emph{Compt. Rend.} {\bf 235},
  557 (1952). Jean-Pierre Vigier, \emph{Compt. Rend.} {\bf 235},
  1107 (1952).
   \bibitem{DeBroglie4} Louis de Broglie, \emph{Compt. Rend.} {\bf 235},
  1345 (1952);{\bf 235},  1453 (1952).
  \bibitem{DeBroglie5} Louis de Broglie, \emph{Compt. Rend.} {\bf 236},
  1453 (1952).
\end{thebibliography}
\end{document}